\documentclass[sigconf]{acmart}
\usepackage{amsfonts}
\usepackage{balance} 
\def\BibTeX{{\rm B\kern-.05em{\sc i\kern-.025em b}\kern-.08emT\kern-.1667em\lower.7ex\hbox{E}\kern-.125emX}}
    
\copyrightyear{2020}
\acmYear{2020}
\setcopyright{acmcopyright}\acmConference[SIGIR '20]{Proceedings of the 43rd International ACM SIGIR Conference on Research and Development in Information Retrieval}{July 25--30, 2020}{Virtual Event, China}
\acmBooktitle{Proceedings of the 43rd International ACM SIGIR Conference on Research and Development in Information Retrieval (SIGIR '20), July 25--30, 2020, Virtual Event, China}
\acmPrice{15.00}
\acmDOI{10.1145/3397271.3401101}
\acmISBN{978-1-4503-8016-4/20/07}

\begin{document}

%
\title{Recommending Podcasts for Cold-Start Users Based on Music Listening and Taste}

%

\author{Zahra Nazari$^{1\ast}$, Christophe Charbuillet$^1$, Johan Pages$^1$, Martin Laurent$^1$, Denis Charrier$^1$,}
\author{Briana Vecchione$^2$, Ben Carterette$^1$}\thanks{$^\ast$ Corresponding author: Zahra Nazari, zahran@spotify.com.}
\affiliation{
\institution{$^1$ Spotify, New York, NY \\ $^2$ Cornell University, Ithaca, NY}
}
%
%
%
%
%
%
%
%
%

%

%
\begin{abstract}

Recommender systems are increasingly used to predict and serve content that aligns with user taste, yet the task of matching new users with relevant content remains a challenge. We consider podcasting to be an emerging medium with rapid growth in adoption, and discuss challenges that arise when applying traditional recommendation approaches to address the cold-start problem. Using music consumption behavior, we examine two main techniques in inferring Spotify users preferences over more than 200k podcasts.
Our results show significant improvements in consumption of up to 50\% for both offline and online experiments. We provide extensive analysis on model performance and examine the degree to which music data as an input source introduces bias in recommendations.  


\end{abstract}

%
%

%

%

%
\maketitle

\par
{\fontsize{8pt}{8pt} \selectfont
\textbf{ACM Reference Format:}\\
Zahra Nazari, Christophe Charbuillet, Johan Pages, Martin Laurent, Denis Charrier, Briana Vecchione, Ben Carterette. 2020. Recommending Podcasts for Cold-Start Users Based on Music Listening and Taste. In 
\textit{Proceedings
of the 43rd International ACM SIGIR Conference on Research and Development in Information Retrieval (SIGIR '20), July 25--30, 2020, Virtual Event, China.}
ACM, NY, NY, USA, 10 pages. https://doi.org/10.1145/3397271.3401101 }

\section{Introduction}



Recommender systems provide an important mode of discovery for users, with more and more content served by platforms via recommendation engines.
Furthermore, platforms are increasingly diversifying, adding videos, podcasts, and other types of media to the content they have traditionally served.
These platforms face the challenge of recommending this new content to users who have never sampled it on that platform and therefore have no history from which to infer their preferences.

Consider podcasts, a relatively new media that has seen rapid growth in recent years: in the US alone, there were {\em three times} as many adults listening to podcasts in 2019 as in 2006, and the number of podcasts available for easy access on platforms like iTunes has grown to an estimated 500,000 in 2018, with over 18 million individual episodes in over 100 different languages. 
Podcasts now comprise an estimated 33\% of audio-only media.
In addition, more and more news and entertainment outlets are turning to podcasts for content delivery.
They are indisputably a major channel for users' consumption online.



Many of the platforms that users turn to for podcasts are also used for other media, most notably music.  Moreover, many of the users of those platforms have long histories listening to music (or consuming other content). 
It may therefore be possible to leverage consumption history for other media to recommend podcasts in the absence of both content analysis and listening history.  
This is known as {\em cross-domain recommendation}. 

The first question we explore in this paper is whether cross-domain recommendation, specifically from music listening history and preferences to podcasts, can be effective.  
It is not clear that this would be the case, as the two mediums differ in many important ways: instrumentation versus spoken word, relatively short songs versus episodes ranging in length from minutes to hours, topical content, and so on.
Thus our first research question is:
\begin{itemize}
    \item[\textbf{RQ1:}] What is the effectiveness of cross-domain models for recommending podcasts to cold-start users based on music preferences?
\end{itemize}
We investigate this by implementing several different models for podcast recommendation on a popular platform for streaming music and podcasts.
Our best models show up to 50\% improvement over simple popularity-based recommendation models in both offline and online tests with real listeners.

We also wish to investigate different representations within the source domain and how they affect the quality of the recommendations.
Users' music listening, taste, and preferences can be represented in a myriad of ways, from simple counts of artist plays to complex embedding models.
Our second research question is:
\begin{itemize}
    \item[\textbf{RQ2:}] What is the most efficient and effective way to represent the user's preferences in the source music domain?
\end{itemize}
We consider three different ways to represent users in the source domain, with increasing fidelity and complexity of computation.
Interestingly, simpler approaches seem to work as well as complex ones.

This leads into our final question, about what we can explain about recommendations in the destination domain. 
\begin{itemize}
    \item[\textbf{RQ3:}] What can we learn about how music preferences influence podcast recommendations? 
\end{itemize}
Specifically, how do our models introduce or propagate bias due to using music as a source?
Simpler representations turn out to be ideal for investigating such questions.

The rest of this paper is organized as follows:  in Section~\ref{section2} we briefly summarize previous work on podcast recommendations, cold-start recommendation, and cross-domain recommendation.
In Section~\ref{section3} we present our models based on transferring representations of music taste to podcasts.
Section~\ref{section4} describes our offline experiments and results, with additional analysis in Section~\ref{section5}.
We conclude in Section~\ref{section6}.

\section{Previous Work}
\label{section2}
Here, we summarize previous work on podcast recommendation, cold-start recommendation, and cross-domain recommendation.

\subsection{Podcast Recommendations}\label{sec:podcast-recs}
While podcast consumption has grown rapidly in recent years, there is comparatively little research on podcast recommendation.
One of the few papers is recent work by Yang et al., which compares the effects of intention-informed recommendations with classic intention-agnostic systems and shows that a recommender can boost a user's aspiration-based consumption~\cite{yang}.

On the volume of previous work available, the Yang et al.\ paper cited above itself notes that only one other prior work seemed to be relevant:  that of Tsagkias et al.\ that ``predicted users' podcast preference using hand-crafted preference indicators''~\cite{tsagkias2010predicting}.

Thus, we claim that this is one of the first studies directly on the topic of podcast recommendation.


\subsection{Cold-start Recommendations}
Podcasts are produced as shows, where each show has a specific theme and releases episodes periodically. Contrary to other domains such as news, where items have a short life span, podcast shows usually last for longer times. This allows systems to gather enough interaction signals for each podcast to match it's audience base. Therefore, the main focus of this paper is to examine solutions for user cold-start rather than item cold-start problem. 
The user cold-start problem~\cite{acilar2009collaborative,schafer2007collaborative} refers to the challenge of making recommendations for users with little to no prior history with the service or medium, or for items that have little historical interaction to draw on.
It is an important problem because cold-start recommendations often form a user's first impression of a recommendation service; if they are poor, the user may be permanently put off.
It remains a difficult challenge, however.

Xu et al.\ categorize cold-start collaborative filtering (CF) approaches into three classes~\cite{xu2017r}:
\begin{enumerate}
\item Onboarding, by which a new user is initially presented with some set of items and asked to express their preferences, often by ratings.  This solution goes back to work by Rashid et al.~\cite{rashid2002getting}.  Onboarding provides necessary data to start making recommendations, but imposes a burden on the user to devote time and energy to this process.

\item Using features of the user~\cite{zhang2014addressing}, the user/item pair, social relationships that can be crawled or scraped from public information~\cite{lin2013addressing}, etc.  Such information can be valuable for making recommendations, but, as we shall see below, is probably not sufficient for making the best possible recommendations.  Furthermore, methods that rely on this information will completely fail when it is not available (due to privacy regulations, omitted data, invisible social media profiles, etc).

 \begin{figure}[t!]
    \centering
    \includegraphics[width=4in]{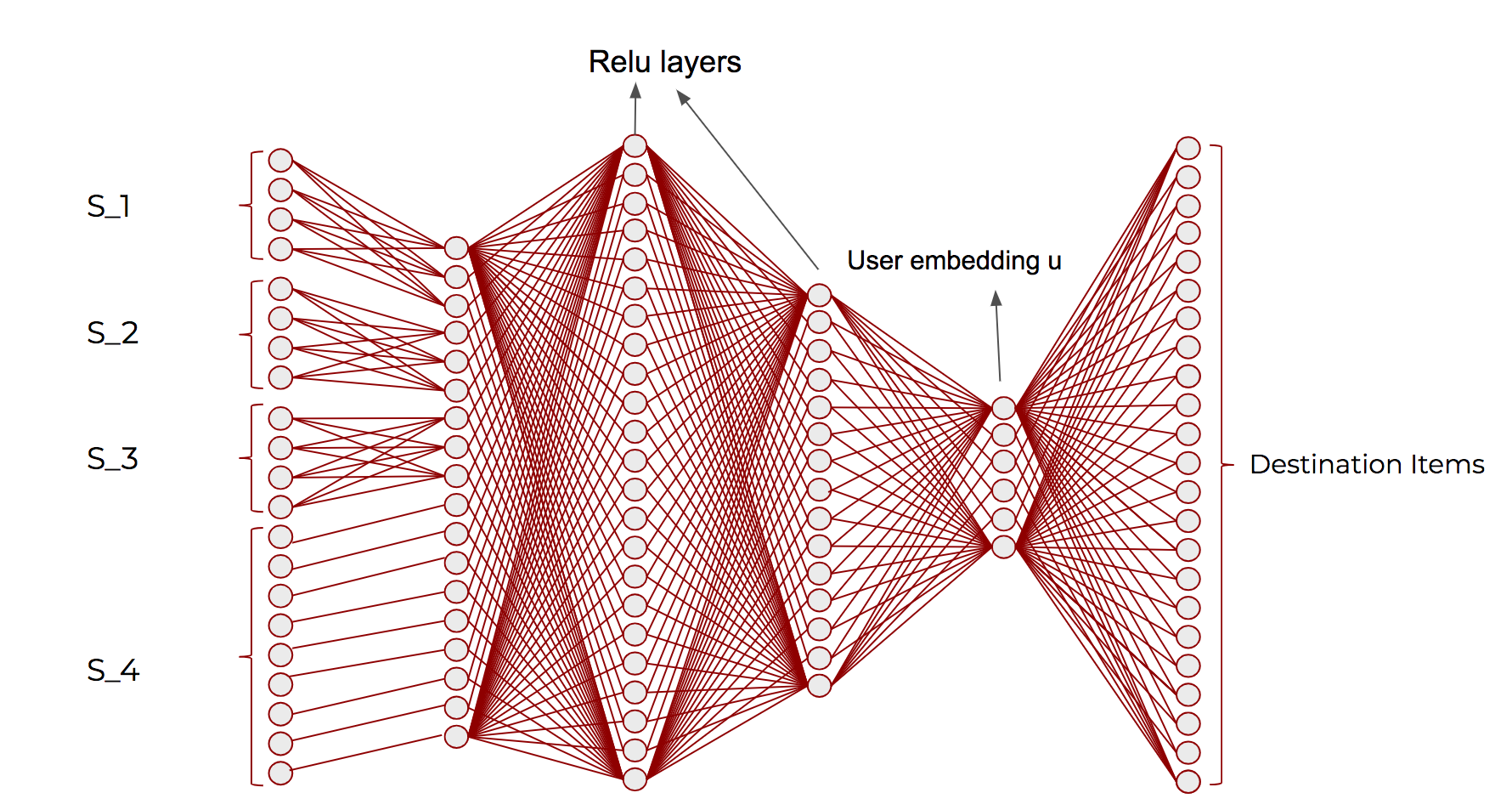}
    \caption{General model for source to destination mapping. The model could include both sparse features (S\_1, S\_2, S\_3) as well as dense ones (S\_4). Sparse features are learnt through the same end-to-end loss function.}
    \label{fig:general-learning}
\end{figure}

\item Dynamically updating user representations in response to user ratings as they appear.  Methods in this class generally try to provide a ``good enough'' recommendation slate for a new user, then quickly update the user representation as the user starts rating items.  Example approaches include the incremental singular value decomposition (iSVD) method~\cite{sarwar2002incremental} and the incremental matrix factorization method~\cite{takacs2008investigation,rendle2008online}.  However, neither method can solve the problem of providing ``good enough'' initial recommendations, nor are they usable for users that do not provide ratings.
\end{enumerate}

In this paper, we introduce a novel approach specifically for podcast recommendation based on users' prior history with other media available on a recommendation platform. Similar to the latter two classes above, our method relies on the availability of historical interaction data, but offers the advantage that nearly all users seeking podcast recommendations will have already used the platform for music. In this way, we are able to utilize a wealth of information about their music preferences which are potentially applicable for podcast recommendation.

\subsection{Cross-domain Recommendations} \label{sec:cross-rec}
Cross-domain recommendation refers to the problem of using representations of users and items in one domain to make recommendations in another domain.  In our case, we show how to use music preferences to make podcast recommendations. 

Fernandez-Tobias et al.~\cite{fernandez2012cross} provide a formal description of the problem in cross-domain recommendation tasks and summarize the existing methods into two categories of content-based and collaborative filtering-based. 

In another work, Fernandez-Tobias et al.~\cite{fernandez2019addressing} introduce a matrix factorization model that jointly exploits user ratings and item metadata to make recommendations on books, movies, and music.  Their approach can provide music recommendations to users who have only rated movies and so forth, but it does rely on having some source of ratings.

Liu et al.~\cite{liu2017transfer} use a neighborhood-based transfer learning to leverage app installation behavior to make recommendations for news.  They show significant improvements over popularity-based news recommendations, though overall performance is quite low.

Mirbakhsh and Ling~\cite{mirbakhsh2015improving} construct a rating matrix that includes available ratings on both of the domains.  They use biased matrix factorization to map this matrix into a lower-dimensional latent space, then use k-means to categorize users and items.

Wongchokprasitti et al.~\cite{wongchokprasitti2015user} propose using a user model that can be shared amongst recommendation services, with a system that maintains the user models based on user interactions with each of these services.  If a single service that provides recommendations in multiple domains fits this description, this is most similar to our proposed approach.

Elkahky et al.~\cite{elkahky2015multi} introduce a content-based approach to cross-domain recommendation, specifically by modeling users across domains.
Finally, Lian et al.~\cite{lian2017cccfnet} use content features to boost their performance in cross-domain recommendations.

\subsection{Our Contribution}
First, as suggested in Section~\ref{sec:podcast-recs}, this work is one of the few on the problem of podcast recommendations in general, and as far as we are aware the first on using music as a source for cross-domain recommendation for the purpose of cold-start podcast recommendation. 
Our results---50\% improvement over popularity-based recommendations in both offline and online tests---show the potential for future research in this area.
Furthermore, our work provides insight into the representation of music tastes for the purpose of podcast recommendation, in particular for explaining recommendations and for investigating bias in them.

\section{Recommending Podcasts based on Music Taste}\label{section3}
Although podcasts and music are both consumed auditorily, they are consumed by users in different ways. Because the most important content within a podcast is spoken, podcasts as a consumption mechanism may be likened more to books than to music. In addition, patterns of consumption vary considerably between music and podcast listening: the number of tracks an active music listener typically listens to during a day is around 20 times more than the similar number for podcasts. However, we consider podcast listening to be a more significant time investment due to the fact that a typical podcast is around 10 times longer than a song. We also find the phenomenon of repeat listening to be a common behavior in music, whereas a single podcast (like a book) is rarely consumed more than once by the same user. 

Ways of modeling and representing music have been studied for a long time, not just by computer scientists but by musicologists, sociologists, and others.
From these studies there are a wide variety of possible features that could potentially be used to represent a user's preferences: audio features, classification features such as genres, co-listening behavior features, co-occurrences on user-made playlists, etc.
Thus we need our model to accept a variety of heterogeneous features in the source domain.

Some of these features would be very sparse.  Take, for example, genre:  some catalogs define more than 3,000 different distinct genres, with any given song falling into only one genre.
Few listeners regularly listen to more than a small fraction of genres, so
a one-hot encoding of genre would result in a very sparse matrix. We require a model that could handle this as well.

Other representations of user taste could be very dense.  We describe one such representation based on playlist co-occurrence below.  Our model also needs to be able to take a dense representation as a feature.

Finally, podcasts, as a new medium, have received comparatively much less study.
We do not know of any work that defines ``standard'' features for podcast representation, nor do we know of many features that correspond to those mentioned above for music.
Rather than develop novel podcast features, we simply use the fact of a user following a podcast as the signal to train to.
This is therefore an extreme multi-class classification problem, with each podcast show being a possible class.

\subsection{Cross-Domain Transfer} 
As we discussed in Section~\ref{sec:cross-rec}, cross-domain recommendation concerns making recommendations in a destination domain $D$ based on representations of users and items in a source domain $S$.

Based on the requirements emerging from the discussion above, and inspired by the CBOW model \cite{mikolov2013distributed} originally proposed for language modeling, we use the framework of recommendations as an extreme multi-class classification task modeled by a multi-layer perceptron (MLP). Previously reported successes of MLPs in various recommendation domains~\cite{cheng2016wide, alashkar2017examples} and in large scale applications~\cite{covington2016deep, chen2017locally, zhang2019deep}, along with their flexibility with heterogeneous sets of features, make this method an appropriate candidate for our problem.

Given a source domain $S$ and a representation $U(S)$ of the user in $S$, we train a neural net that maps a user to a distribution over items in the destination domain $D$.
A representation of sparse features (e.g., $_1$, $S_2$ and $S_3$ in Figure 1) could be learnt through an end-to-end training, while dense features (e.g., $S_4$ in Figure 1) could be memorized through the network and concatenated to the final user representation at any stage.

We use a softmax classifier to minimize the cross entropy loss for the true label and the negative samples. Let $i \in D$ be the label, and $d_i$ be an $N$-dimensional vector representation of $i$.  A user's preferences, $U(S)$, are represented as an $N$-dimensional vector $u$. 
\begin{align*}
P(i | U(S)) = \frac{e^{d_i u}}{\sum_{j\in D}e^{d_j u}}
\end{align*}

Optimizing the network this way would result in the dense vector $u\in \mathbb{R}^N$ being closer to the item $i$'s vector as the weights of the node $i$ in softmax classifier: $d_i\in \mathbb{R}^N$. 

Because each podcast show is a class, we need to be able to handle a large number of classes in training.
We use importance sampling, a negative sampling approach proposed by Jean et al.~\cite{jean2014using} to have the model converge in efficient time.
The loss function is then calculated as: 
$$J_\theta = -\sum_{i\in D} [\log\sigma(ud_i)+\sum_{j=1}^k\log\sigma(-u{d_{ij}})]$$
where $k$ is the number of sampled negatives, $d_{ij}\in \mathbb{R}^N$ is the vector for the $j$th negative class sampled for label $i$ and $\sigma(x) = \frac{1}{1+\exp(-x)}$

Given this modeling framework, we now consider music as the source domain, with $S_1, S_2, S_3, S_4$ in Figure~\ref{fig:general-learning} illustrating various sparse and dense features of a user's listening habits.  We use podcasts as the destination.  
Training the network learns a mapping from music to podcast preferences.

\subsection{Representing Music Taste}
In order to efficiently represent a user's source profile for cross-domain recommendation, we consider a range of approaches, starting from simple demographic-based heuristics to more complex pre-trained representations of music taste.
These approaches are based on features that are available on a popular music streaming service.

\subsubsection{Demographics}
The simplest user representation uses basic demographic information about a user to infer their music taste. This information may include country, age, and gender, all of which a user can choose to self-report on many platforms.

\subsubsection{Metadata}
We can generate a better representation of music preferences using music metadata information such as artist and genre. We represent users as a simple aggregation over their music consumption metadata. For example, users could be represented by their top listened artists and genres. We also use manual annotations for three levels of genres:  meta-genre, a high-level music category such as ``folk'' or ``rap''; genre, which includes more specific types of music such as ``blues'' or ``hip hop''; and micro-genre, which defines niche subgenres and includes labels such as ``Texas blues'' or ``East coast hip hop''.
Overall, our data has more than 1.3 million distinct artists, 40 genres and 3855 micro-genres with which to represent users.

\subsubsection{Latent Representation}
In a more complex context, we can take a modelling approach and use contextual and collaborative information to embed users into a latent representation in their source domain. Either CF, content filtering, or a hybrid approach could be sufficient to calculate these representations. In order to obtain this, we use a rich dataset of user-created playlists.


First, we represent all tracks in our catalog using a high-dimensional vector trained based on co-occurrence of tracks in a playlist. The vectors were trained using an embedding model inspired by word2vec ~\cite{pennington2014glove} which was originally introduced for learning word embeddings. More specifically, we used the Word2vec Skip-gram model, a shallow neural network with a single hidden layer that takes a track as input and predicts the rest of the tracks in the playlist. Let's consider a playlist containing $song_A$, $song_B$,and $song_C$ as an example. The shallow neural network is then trained using $song_B$ as the input with $song_A$ and $song_C$ as the target. The network exploits the co-occurrence of songs in a playlist to learn a high dimensional vector representation for each track, where similar tracks will have weights that are closer together than tracks that are unrelated.
A user representation is then calculated as recency-based weighted average over their listening behavior. The resulting output is a dense vector that could directly be used as the user embedding.

\section{Experiments and Results}
\label{section4}
We now turn to experiments.  In this section, we describe a set of offline experiments and one online experiment for cross-domain podcast recommendations based on music. 

\subsection{Data}
Our data consists of 17 million users who follow at least one podcast on Spotify. On average, each user follows 2.9 podcasts. In addition to the podcasts they follow, each user has a list of songs (referred to as ``tracks'') that they have listened to on the platform along with annotated information for each track, including artist and genre.

Specifically, the data includes:
\subsubsection{Demographics} For each user, we have access to self-reported information on country, gender and age. 
    In this group of users, 17 different self-reported countries of origin are reported.
    Gender is classified and reported as male, female, neutral or none. Users' age is partitioned into eight buckets including ``unknown''.

\subsubsection{Metadata}
For each user, we have a list of tracks that user has listened to on the platform during the previous 90 days. Each track is associated with various metadata including artists, genres, meta-genres, and micro-genres. These genres are manually tagged by music experts.

\subsubsection{Music Taste Embeddings}  In addition to the user's music behavior, we use a dataset of nearly 700 million playlists curated by users. Each playlist consists of 10 tracks on average, each of which has been put together by one or more users on platform.
These playlists are used in a CF approach using word2vec to create 40-dimensional vector representations of tracks as described above.
A user embedding can be calculated simply as a recency-weighted average over track embeddings.

Note that these embeddings are very costly to compute.  In practice it takes several hours of computation time, and they need to be recomputed frequently due to changes in the catalog and in user playlists.  Furthermore, they are opaque: they provide no intuition for what a value on any given dimension represents, and thus do not explain anything about recommendations.

\subsubsection{Podcast Interests} Spotify gives users the option to follow their favorite podcast shows. For each user we have a list of podcast shows that they follow.


\subsection{Offline Experiments}
\begin{figure*}[t!]
    \centering
    \includegraphics[width=4in]{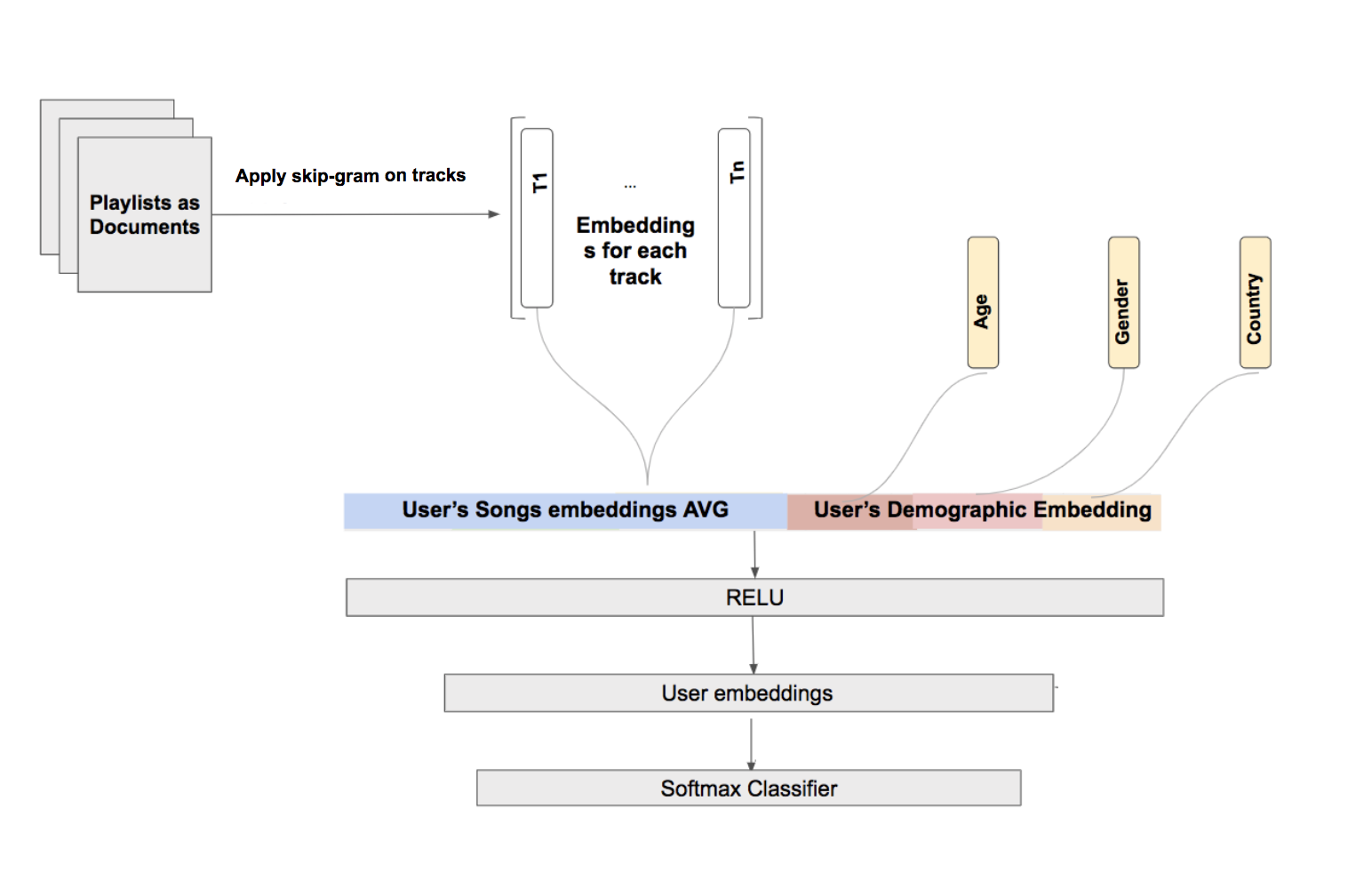}
    \caption{ Learning from Music Representation}
    \label{fig:item-model}
\end{figure*}

\subsubsection{Training and test splits}

We split the data by user ID, with about 200,000 unique users in the test set and the rest in the training set.
The training set contains one instance for each individual user/podcast follow, so if one user follows 10 podcasts there will be 10 separate instances for that user reflecting listening history, demographics, and an indication that they follow a specific podcast.
By comparison, the test set contains one instance per individual user. After training, models rank a set of podcasts for each user in the test set, and that ranking is evaluated by retrieval metrics such as precision and recall (see Section~\ref{metrics} below).

\subsubsection{Models}\label{models}

We compare nine different models for podcast recommendation.  Two are baselines that rank popular podcasts, three are cross-domain CF models, and the remaining four combine CF models with demographic data.

\begin{enumerate}
\item \textbf{Popularity by country}: This is a simple popularity-based model that recommends the most popular podcasts in the user's country of origin.  
\item \textbf{Popularity by country and demographic}:  This refines the previous model to recommend popular podcasts for the user's country of origin as well as demographic (gender and age group). 

\item \textbf{Cross-domain CF using logistic regression}: Since we define our problem as a pure cold-start task, we use popularity per demographic as an optimal baseline. In order to motivate the need for an MLP approach, we additionally evaluate a logistic regression model as a second baseline. This model uses as inputs the embeddings described above to predict a podcast follow for each user. In order to ensure fair comparison, we used the same softmax layer and negative sampling approach used by all other models in the study.

\item \textbf{Cross-domain CF}: This approach introduces our first MLP based model, which uses the embeddings described above and podcast follows in the training data to fit podcast vectors. Podcast recommendations are produced from the $k$ nearest neighbors to the user embedding in the test set.

\item \textbf{Cross-domain CF with metadata}:  This model uses a user's top-$m$ listened artists, genres, meta-genres, and micro-genres to fit podcast vectors.  

\item \textbf{Demographics + cross-domain CF}:  This model uses the user embeddings along with vectors representing demographic features to fit podcast vectors.  Figure~\ref{fig:item-model} shows the complete model.
\item \textbf{Demographics + cross-domain metadata CF}:  Similarly, this model uses top-$m$ entities in metadata categories along with demographic features.
\item \textbf{Cross-domain CF + metadata}:  This model combines the user embedding with the metadata categories, but does not include demographic features.
\item \textbf{Demographics + all music data}:  This model combines the user embedding, metadata, and demographic features into a single model.
\end{enumerate}

\subsubsection{Metrics}\label{metrics}

As mentioned above, each model produces a ranking of podcasts for each user in the test set. Given these rankings, the gold standard is whether that user actually follows the podcasts suggested in the ranking. This is evaluated using the standard retrieval metrics precision, recall, and nDCG.

Specifically, we look at precision at rank 1, defined as the proportion of user requests in the test set that are recommended a podcast they follow at rank 1. We consider the limitation that this metric may be too coarse, as it can only be zero or one.
Because of this, we also look at precision at rank 10 to investigate the proportion of podcasts a user follows that we are able to correctly recommend. Please note that since on average each user follows 2.9 podcasts, the upper bound of precision at rank 10 is about 0.29.

nDCG weights each item by a function of the rank at which it appears. This is important for the context of podcast recommendations because users typically see only a few recommendations in the interface and must scroll or click to see more. The effort this takes means that fewer users see lower-ranked recommendations, so relevant recommendations not initially seen by the user are down-weighted by rank to capture this. nDCG discounts relevance by $\log(\text{rank}+1)$.

\subsubsection{Hyper-parameter tuning}
We use a portion of our training data as the validation set to tune a number of hyper-parameters: the embedding size for each of the input features,  the number of fully connected layers,  the final layer size for user embeddings, and the number of negative samples. For demographic features, we experimented with 5, 10 and 15 as the embedding size, and found using 10 for each feature worked best. Increasing the hidden fully connected layers from one to two increased the accuracy, but having three layers did not improve the results. For user embedding layer size, we observed that increasing the size from 20 to 40 improved the results, while any size larger than 40 dropped the accuracy. For number of nodes in each hidden layer, we ran experiments with 128, 256, 512, 1024. Although increasing the number of nodes in each hidden layer resulted in slightly better results, we decided to use 512 to keep our training time under 24 hours. The last parameter was the number of negative samples, where we experimented with 256, 512, 1024, 2048, 4096, and found 2048 to perform best.

\begin{table*}[t!]
\centering
 \begin{tabular}{|l || c c c c c |} 
 \hline
 \textbf{Model}  &	\textbf{ndcg@10}	& \textbf{ndcg@50} &	\textbf{precision@1} &	\textbf{precision@10}&	\textbf{recall@10} \\ [0.5ex] 
 \hline\hline
\textbf{country popularity} &	0.12256 &	0.17641	& 0.07722 &	0.04188	& 0.19027 \\ 
 \textbf{country + demo popularity} &	0.15310 &	0.20817&	0.10447&	0.04814	&0.21850 \\
 \hline
 \textbf{cross-domain CF Logistic Regression} & 0.16610	& 0.21690 &	0.11779 &	0.04974 &	0.23268 \\
\textbf{cross-domain CF} &	0.19029	& 0.24605 &	0.13939 &	0.05680 &	0.26231 \\
\textbf{cross-domain metadata CF}&	0.21212&	0.26933	&0.16056&	0.06274	&0.28710 \\ 
\hline
 \textbf{demo + CF} &	0.20568&	0.26399&	0.15197&	0.06129&	0.28199 \\
\textbf{CF + metadata}	&0.21327	&0.27058&	0.16173	&0.06294&	0.28877 \\
\textbf{demo + metadata}	&0.21938$^a$ &	0.27762	& \textbf{0.16631}$^b$ &	0.06486$^c$	&0.29658$^d$ \\
\textbf{demo + CF + metadata} &	\textbf{0.22009}$^a$	&\textbf{0.27866}&	0.16623$^b$	&\textbf{0.06503}$^c$ &	\textbf{0.29763}$^d$\\
[1ex] 
 \hline
 \end{tabular}
 \caption{Results of offline experiments comparing podcast recommendation using models based on popularity, CF, CF in metadata, and combinations of the latter two with demographic information. Most differences are statistically significant.  Pairs marked $^a$, $^b$, $^c$, and $^d$ are {\em not} significantly different.}
 \label{table:offline}
\end{table*}

\subsection{Results}
Table~\ref{table:offline} summarizes offline results for all eight models, with models grouped as described in Section~\ref{models}. We used Tukey's Honest Significant Differences to perform statistical significance testing with correction for multiple comparisons; all differences are statistically significant except for those noted by superscripts.

It is clear that all of our music cross-domain models perform significantly better than popularity-based models, even when those models are tuned to country of origin and demographic.  This directly answers our first research question, that indeed music-based cross-domain models are effective in recommending podcasts.  

The demographic+country popularity model is 20\% worse than the cross-domain CF model, while the country-only popularity model is 36\% worse.  This trend is seen across all measures, and while the \% differences vary, it is always the case that both popularity models are substantially worse than the CF model. Moreover, the country popularity model is substantially worse than the country+demographic popularity model. All of these differences are statistically significant. 

Another insight from our evaluations is that although a logistic regression model over user embedding vectors improves the popularity baselines, the results motivate the need for having non-linear models to have better results. This goal is achieved using multi-layer perceptrons in the next sets of models. 

We continue to compare other models based on CF, starting with the two MLP based cross-domain models.  The metadata model is the clear winner, outperforming the CF model by 11\% on nDCG@10 and similar margins on other metrics.  Though both of these methods are based on CF, the metadata model uses more granular information about music tastes. This appears to benefit podcast recommendation.  Combining these two methods gives superior results than either do independently. The CF + metadata model shows a 12\% improvement over the base CF model, though less than 1\% over the metadata model, suggesting that the metadata model captures most of what is necessary about music tastes to recommend podcasts, without the need to perform the word2vec algorithm on the playlist data. 

Adding demographic information to these two models continues to boost effectiveness, though at a lesser rate.  For the base CF model, the gain in nDCG@10 is about 8\%; for the metadata model, the gain is only about 3\%.  We see similar gains across other metrics.  

Finally, the best results in an absolute sense are achieved when CF is combined with metadata-based CF and demographic information.  
Note, however, that adding the CF does not present  significant gain in effectiveness---only the difference in nDCG@50 is statistically significant.
This suggests that the complexity and computation time of the vector representation is not worth it.
Metadata-based representations are far easier to compute and provide essentially the same benefit. 
This helps answer our second research question regarding efficient and effective ways to represent user's preferences.

\subsection{Online Experiments}

The ultimate objective of this work is to be able to match new users to relevant podcasts without any prior knowledge about their podcast affinities. Therefore, we implemented our best performing model---the demographic + cross-domain CF + metadata model---in production on Spotify and compared it with our best performing baseline. 
The test was performed on 800,000 users with no history of podcast consumption on the platform.
The users were randomly split into two groups:  the control group received a slate of 10 podcast recommendations generated by the country + demographic popularity model, and the treatment group received a slate of 10 podcast recommendations generated by the demographic + cross-domain CF + metadata model.
Over the course of one week, we measured podcast consumption in minutes and the number of shows newly followed by these users.

\begin{figure}[t!]
    \centering
    \includegraphics[width=2.5in]{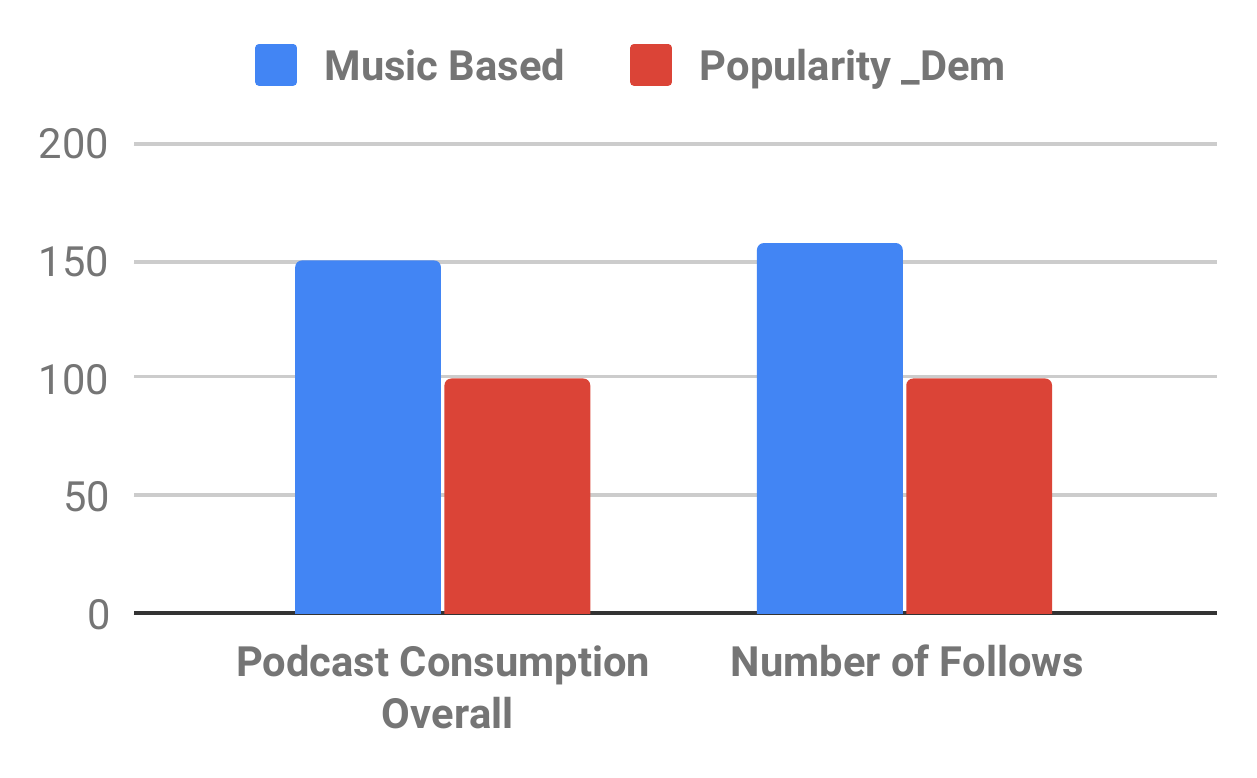}
    \caption{ Results from an online experiment comparing the country + demographic popularity model to the music-based cross-domain CF + metadata model. Note that results are scaled such that the baseline is 100\% in order to directly reflect the percent improvement.}
    \label{fig:online}
\end{figure}

Figure~\ref{fig:online}  shows the results. The group exposed to music-based recommendations listened to nearly 50\% more podcast minutes than the control group and followed over 50\% more shows.
This provides validation of our offline experiment and offers striking confirmation of our first research question:  that music tastes can predict podcast interest.

\section{Analysis}
\label{section5}
In this section, we first examine models' effectiveness across various user cohorts. Then, we present diversity-related effects of using music as a training source for prediction. Lastly, we describe several anecdotal examples where newer models perform better than the baseline. These analyses shed insight into the model's learned associations between music tastes and podcasts, our third research question.

\subsection{Model Performance in User Cohorts}
In this section, we analyze model performance across the following four dimensions of interest:
\begin{itemize}
    \item \textbf{Country of origin}: One of the main motivations of this work is to be able to recommend relevant spoken content without going through the costly process of understanding content across a variety of languages. Therefore, performance across different countries and languages is an important target. Figure~\ref{fig:countries} shows the podcast follow results for our largest user base countries. We improve the performance by at least 30\% across all countries.
    \begin{figure}[t!]
    \centering
    \includegraphics[width=3in]{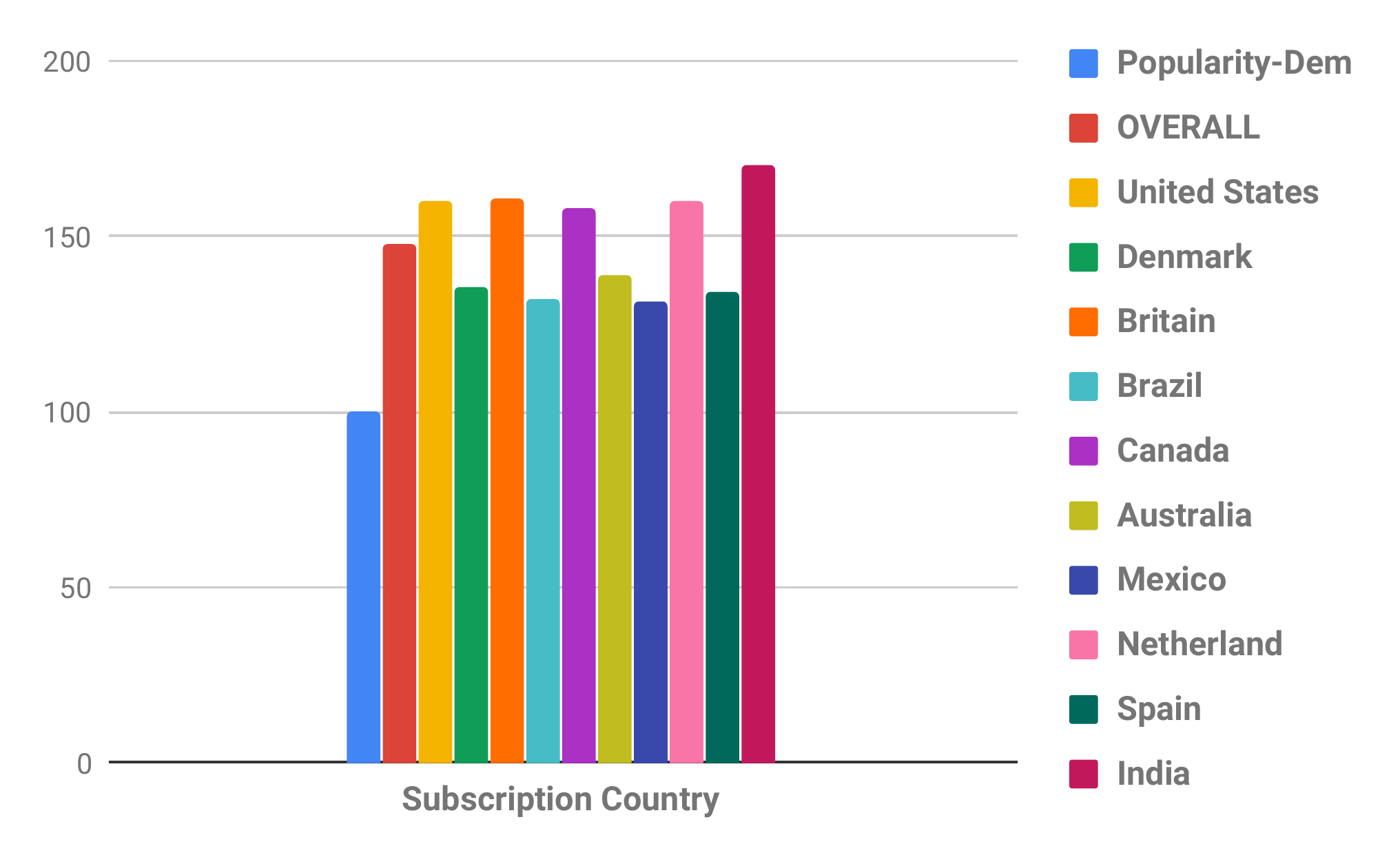}
    \caption{Podcast follow prediction performance of our best model compared to popularity baseline across countries. The blue and red bars are overall results for this group of countries. Each subsequent bar shows the relative improvement over the country + demographic popularity model for that country.  }
    \label{fig:countries}
\end{figure}

    \item \textbf{Time on platform}: The amount of music listening history required for high model performance is important to accurately recommend podcasts of interest. The age of each user's account is a good proxy for this; Figure~\ref{fig:account_age} shows that the best performing model is able to outperform the popularity-based model using less than a week of music consumption information. In fact, users do not need to have long listening histories in order to receive good podcast recommendations.
    \begin{figure}[t!]
    \centering
    \includegraphics[width=2.5in]{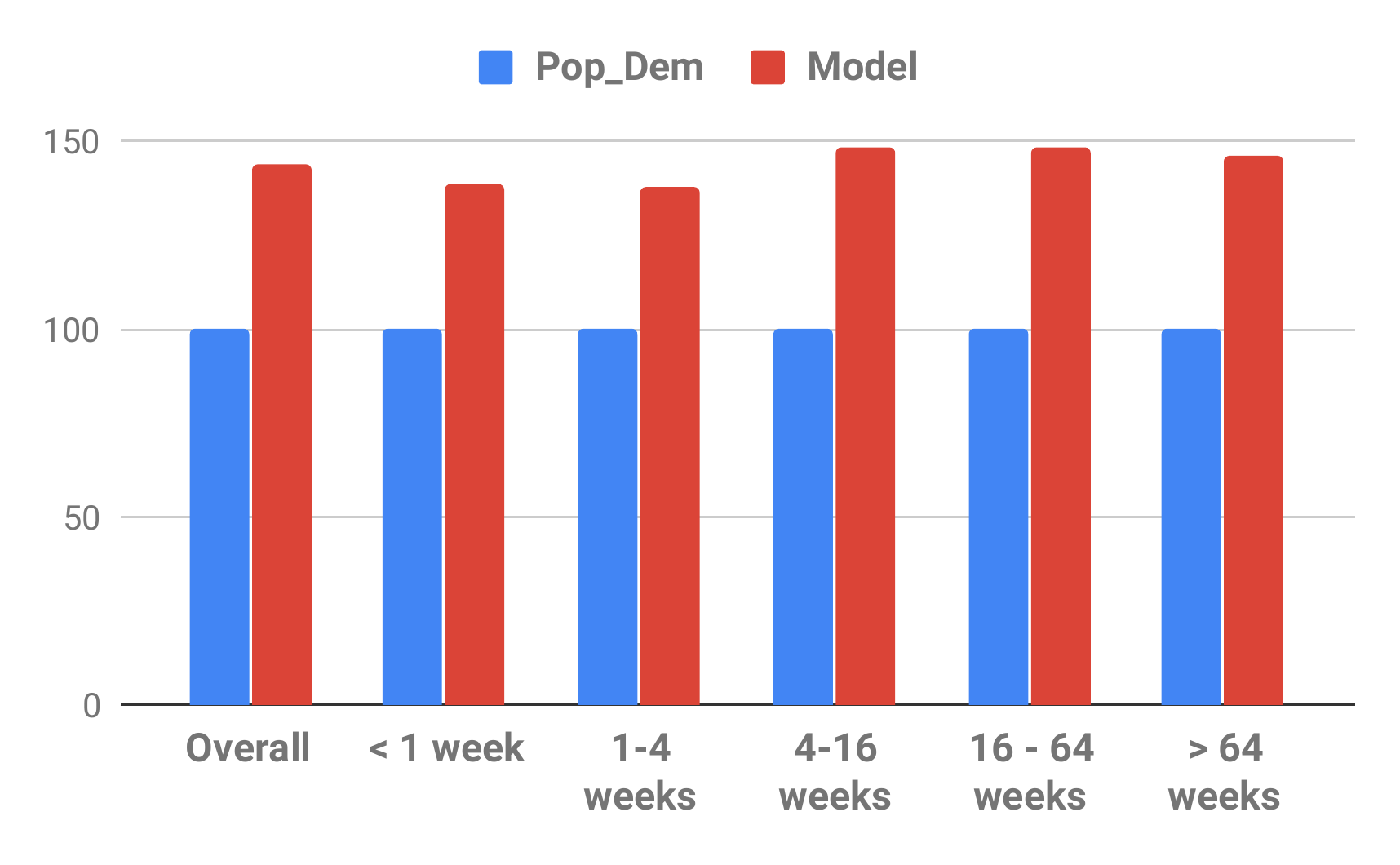}
    \caption{ Podcast follow prediction performance of our best model compared to popularity baseline across account ages. }
    \label{fig:account_age}
\end{figure}

    \item \textbf{Age bucket}: We are interested in evaluating performance across all age ranges in order to see if some age ranges were being underserved by the model. Figure~\ref{fig:age_bucket} shows that our model performs best in the [25-29] age bucket with a 50\% improvement over the baseline, but performs well across all age buckets.
\begin{figure}[t!]
    \centering
    \includegraphics[width=2.5in]{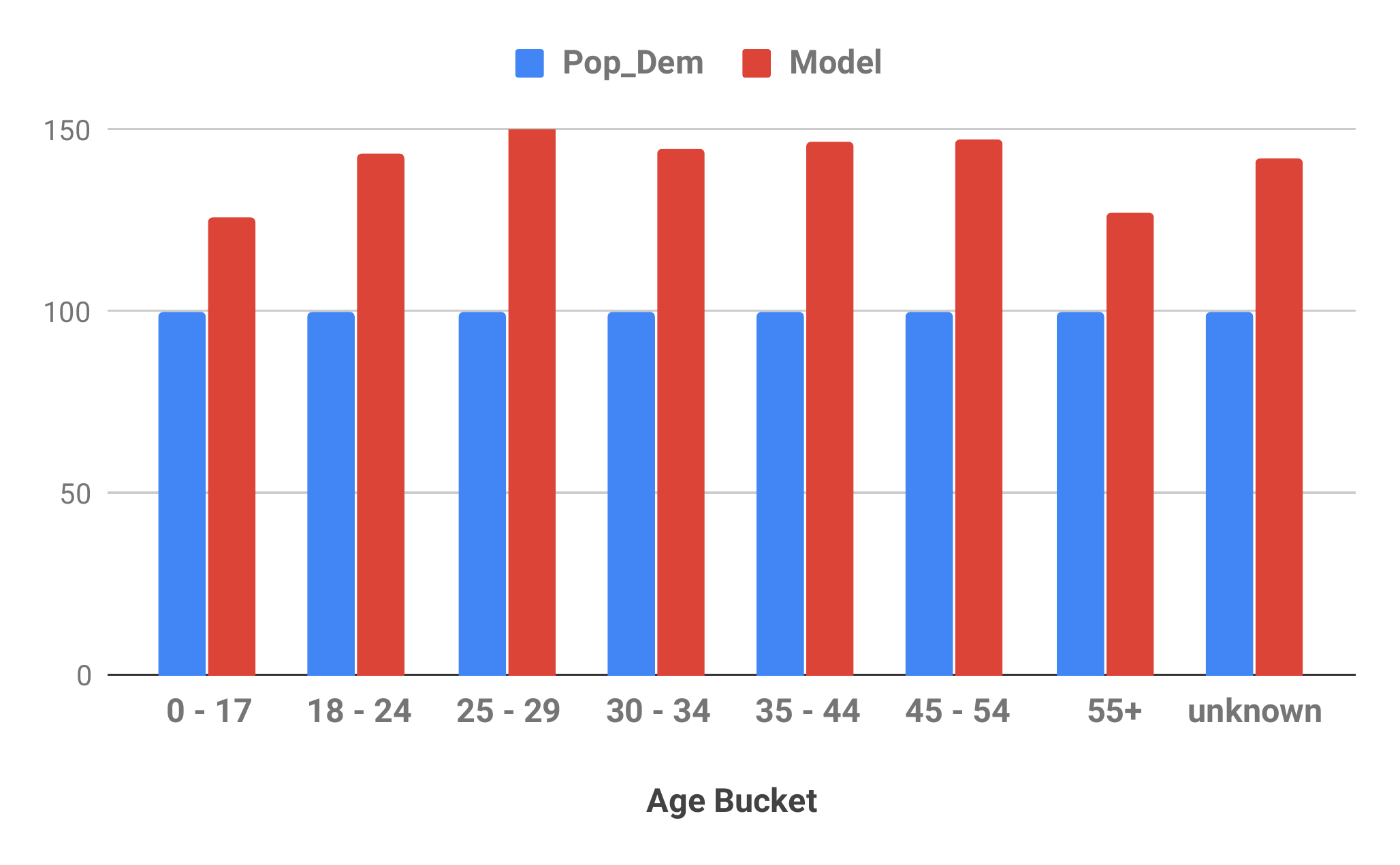}
    \caption{ Podcast follow prediction performance of our best model results compared to popularity baseline across user age buckets. }
    \label{fig:age_bucket}
\end{figure}

    \item \textbf{Gender} Finally, our model's performance for self-reported genders are shown in Figure~\ref{fig:gender}. We did not observe any significant difference between male and female categories.
\begin{figure}[t!]
    \centering
    \includegraphics[width=2.5in]{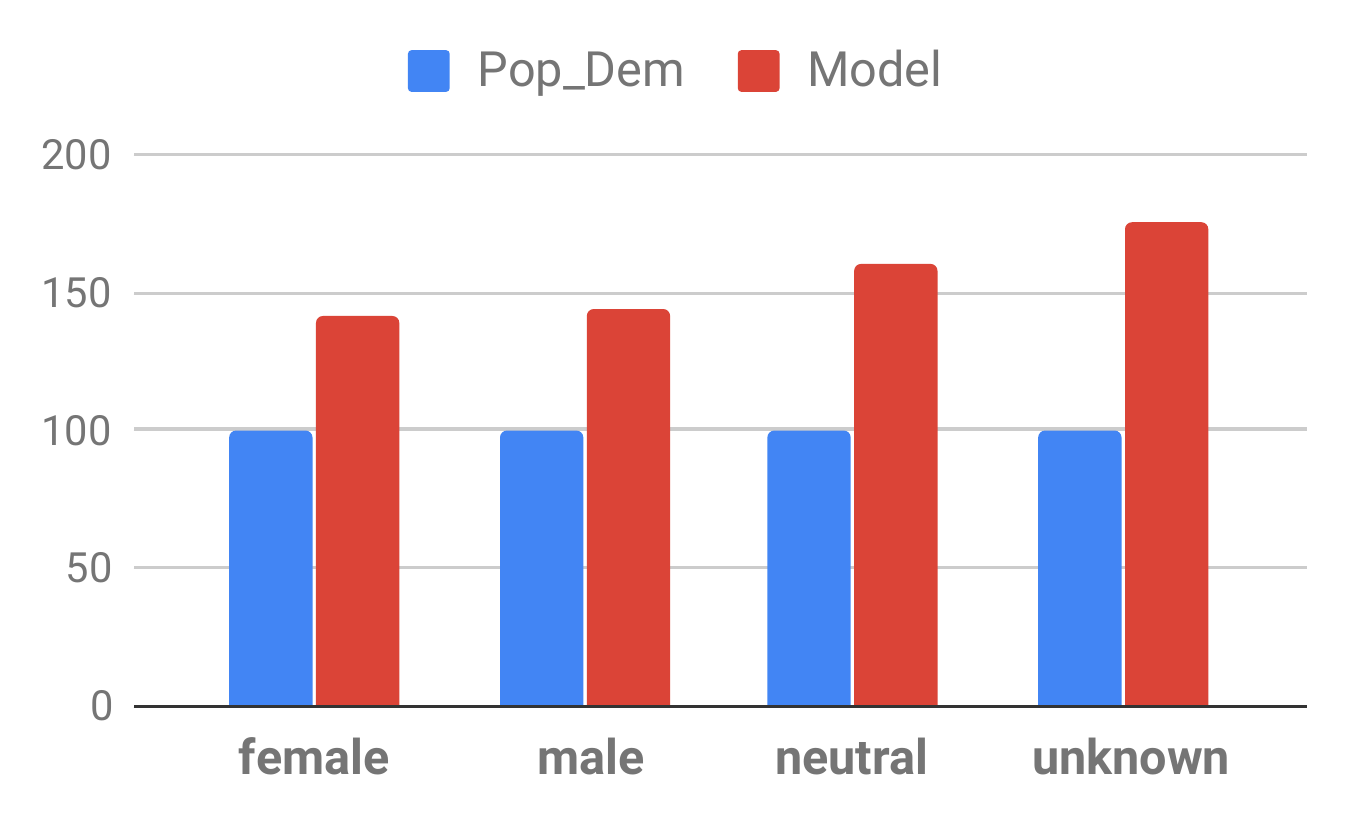}
    \caption{ Podcast follow prediction performance of our best model results compared to popularity baseline across user genders. }
    \label{fig:gender}
\end{figure}

 \begin{figure}[t!]
    \includegraphics[width=3.4in]{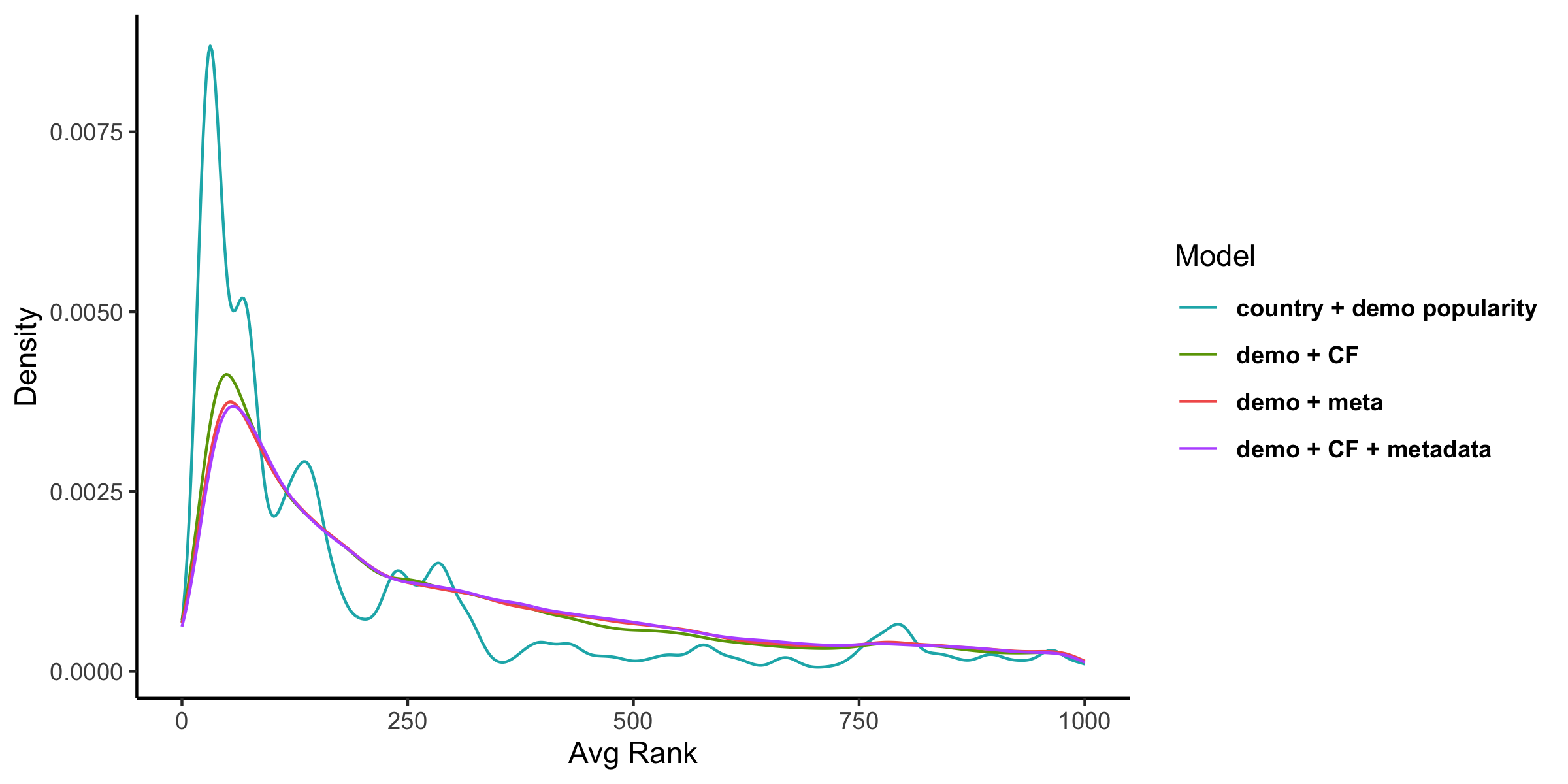}
    \caption{Popularity distribution for podcast recommendations across models. }
    \label{fig:popularity}
\end{figure}

\end{itemize}

In addition to the positive results shown here, this also shows that a demographic representation is suitable for asking further questions about the effectiveness of recommendations.

\subsection{Diversity Analyses}
 We evaluated various dimensions of recommendations produced by a subset of models in order to understand the extent to which our input data may propagate bias into our recommendations. Specifically, we assess the models for instances of \textit{popularity bias}, which occurs when recommenders prioritize popular items much more highly than other items in a long tail distribution regardless of user taste \cite{park2008long}. The goal here is to assess whether popularity bias is necessary to achieve better performance.

We explored the extent of heterogeneous recommendations according to their overall popularity rank for models: ``demo + CF'', ``demo + metadata'', and ``demo + CF + metadata''. We compare these recommendations against the baseline model, ``country + demo popularity.'' Figure~\ref{fig:popularity} shows the average rank of top 10 podcast recommendations across models for a random sample of users. The baseline shows a long tail distribution skewed towards recommendations from top-ranked positions, while models with improved performance show considerably less homogeneity. It is notable to mention that our best performing model suggests the least homogeneous set of recommendations on average. This implies that we are able to simultaneously show users more niche and accurately-tailored content while also improving performance.
\begin{figure}[t!]
    \includegraphics[width=3.5in]{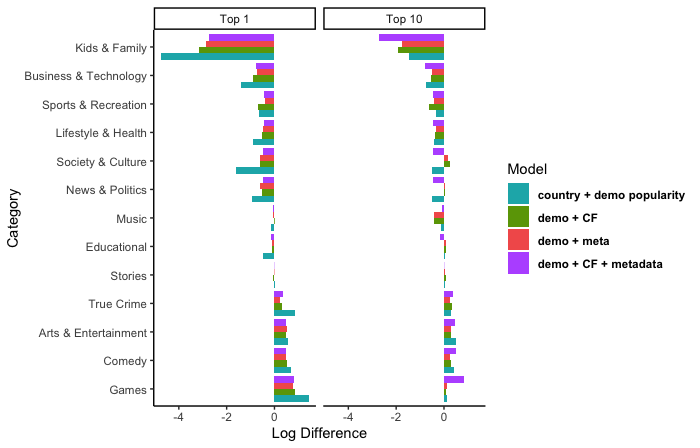}
    \caption{Podcast category differences compared to user follows across models for top 1 and top 10 recommendations.}
    \label{fig:categories}
\end{figure}
    
\begin{figure*}[t!]
    \centering
    \includegraphics[width=5.2in]{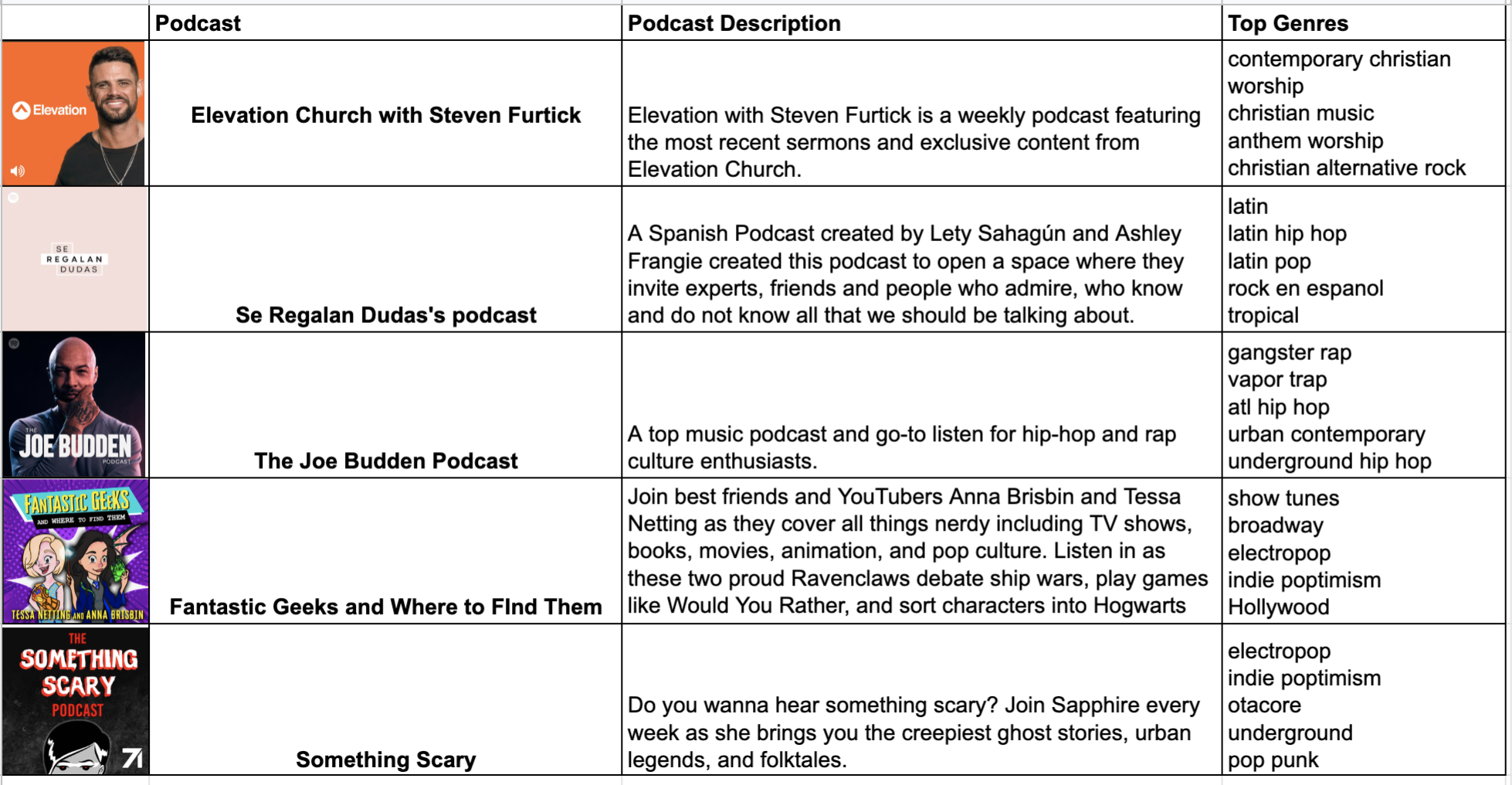}
    \caption{ Some examples of podcasts that online users followed when recommended by the demographic + cross-domain CF + metadata model, along with the top genres preferred by those users.}
    \label{fig:examples}
\end{figure*}

Next, we assessed the extent to which our models introduce \textit{behavioral bias}, which we define as the difference between organic user consumption behavior and model recommendations. Because models are trained using music listening data, we wanted to explore whether models bias users to consume higher rates of podcasts annotated with a music label. We use annotated podcast categories as our dimension to approximate organic behavior. Since each podcast may be labeled with up to ten categories in no order of importance, we weight categories and aggregate the percentage of recommendations each category comprises. We calculate the log difference for each podcast category, which is defined as $\log(x_2/x_1)$ where $x_1$ represents the percent of organic user follows for each category and $x_2$ represents the percent each category is recommended across models. Figure~\ref{fig:categories} shows the difference between categories that users follow organically on the platform compared to the top $n$ recommendations suggested across models. Educational content is the most followed category while content labeled as kids \& family is followed the least. We see that more popular categories tend to show smaller differences while less popular categories are recommended less often to users. The figure shows that our best performing model is able to reflect organic user behavior as well as serve users with a more equal distribution of podcast categories compared to the popularity baseline. Overall, the models perform in accordance with organic user behavior and do not show immediate suggestions of behavioral bias related to category.

\subsection{Qualitative Analyses}
We examined some individual podcast metadata in order to gain insight about the associations between music taste and podcasts being generated by our models. Specifically, we selected for US users due to our better understanding around both content as well as accuracy of genre annotations.
Figure~\ref{fig:examples} shows some podcast examples and their descriptions. These podcasts were selected at random from the pool of users for whom our model was correctly recommending podcasts when the baseline model was failing to do so. For the sake of clarity, the top five most popular genres are removed from the list. 

Some of these examples show interesting connections between podcasts a user subscribes to and music genres they listen to. For example, \textit{Elevation Church with Steven Furtick} is a religious podcast, so users who are correctly receiving this podcast as a recommendation have genre tastes including contemporary Christian, worship, Christian music, anthem worship, and Christian alternative rock. Although language is not an explicit input to these models, language seems to be learned by the model, e.g.\ Spanish podcasts are recommended to users who have an affinity for Latin music but may not necessarily be located in a Spanish-speaking country. We also noticed that a strong music-related theme in a podcast could be seen in the type of music their subscribers are listening to. Although we show genre information in Figure \ref{fig:examples} for the sake of explainability, the models that do not use genre information, i.e.\ demo + CF, correspond with generally equivalent results. This leads us to believe that our models indirectly pick up music genres and do not need fine-tuned annotations to perform well.

\balance
\section{Conclusion}
\label{section6}
In this paper, we examined the viability of using past music listening and representations of musical taste as a source domain to recommend items in the target domain of podcasts at Spotify. Due to differences in media types and consumption patterns, it appears there should be little reason for music taste to be able to accurately predict podcast listening. Yet, even with no prior information about users' interest in podcasts, we are able to increase both minutes listened and podcasts followed by 50\% in online experiments, answering our first research question positively and decisively.

We also compared two main approaches in leveraging rich interaction signals in a source domain for recommendations in a destination domain. We showed that although an approach that includes both an item-based CF model and meta-data signals would do best, the meta-data based method alone could achieve comparable results that are both more explainable and more efficient. 
This answers are second research question about representations.

Our analysis shows that these models work well in all countries, across all ages, for all genders, and do not require a great deal of listening history to perform well for podcast recommendation. Further, these models are able to decrease popularity bias without over-representing music-related content. This analysis is suggestive towards our third research question regarding the effect of music preferences on podcast recommendations, though we believe there is still more work to be done.

For future work, we plan to test this model for warm-start users as well. In this task, we expect to recommend new podcasts to users who already follow one. Additionally, we plan on incorporating podcast features as well, such as language, distributer, and audio embeddings. It seems intuitive that both of these directions will significantly impact user engagement with podcasts.
We also note in Section~\ref{models} that as a side-effect of our models, we obtain vector representations of artists and genres which encode the relationship between a user's taste in music, their favorite artists and genres, and the podcasts they follow. We plan to investigate whether these vectors could be useful for larger-scale cross-domain recommendation as well as recommending artists, shows, and more.

\clearpage
\bibliographystyle{ACM-Reference-Format}
\balance
\bibliography{sample-base}

\end{document}